# Million-*Q* Free Space Meta-Optical Resonator at Visible Wavelengths

Jie Fang[1,#], Rui Chen[1,#], David Sharp[2,#], Enrico M. Renzi[3,4], Arnab Manna[2], Abhinav Kala[1], Sander A. Mann[3], Kan Yao[5], Christopher Munley[2], Hannah Rarick[2], Andrew Tang[1], Sinabu Pumulo[6], Yuebing Zheng[5], Vinod M. Menon[4,7], Andrea Alù[3,4], Arka Majumdar[1,2]

[1] Department of Electrical and Computer Engineering, University of Washington, Seattle, WA 98195, USA.

[2] Department of Physics, University of Washington, Seattle, WA 98195, USA.

[3] Photonics Initiative, Advanced Science Research Center, City University of New York, New York, NY 10031, USA.

[4] Physics Program, Graduate Center, City University of New York, New York, NY 10016, USA.

[5] Walker Department of Mechanical Engineering and Texas Materials Institute, The University of Texas at Austin; Austin, TX 78712, USA.

[6] Department of Materials Science and Engineering, University of Washington, Seattle, WA 98195, USA.

[7] Department of Physics, City College of New York, New York, NY 10031, USA.

[#] These authors contributed equally to this work: Jie Fang, Rui Chen, David Sharp.

**Abstract**

High-quality ($Q$)-factor optical resonators with extreme temporal coherence are of both technological and fundamental importance in optical metrology[1,2], continuous-wave lasing[3–5], and semiconductor quantum optics[6–8]. Despite extensive efforts in designing high-$Q$ resonators across different spectral regimes, the experimental realization of very large $Q$-factors at visible wavelengths remains challenging due to the small feature size that is sensitive to fabrication imperfections[9–12], and thus is typically implemented in integrated photonics[1–4,9,10]. In the pursuit of free-space optics with the benefits of large space-bandwidth product and massive parallel operations[13,14], here we design and fabricate a visible-wavelength etch-free metasurface with minimized fabrication defects and experimentally demonstrate a million-scale ultrahigh-$Q$ resonance. A new laser-scanning momentum-space-resolved spectroscopy technique with extremely high spectral and angular resolution is developed to characterize the record-high $Q$-factor as well as the dispersion of the million-$Q$ resonance in free space. By integrating monolayer WSe$_2$ into our ultrahigh-$Q$ meta-resonator, we further demonstrate laser-like highly unidirectional and narrow-linewidth exciton emission, albeit without any operating power density threshold. Under continuous-wave laser pumping, we observe pump-power-dependent linewidth narrowing at room temperature, indicating the potential of our meta-optics platform in controlling coherent quantum light-sources. Our result also holds great promise for applications like optical sensing, spectral filtering, and few-photon nonlinear optics.

High-quality ($Q$)-factor optical resonators with ultranarrow spectral linewidth ($\Gamma_\omega = \omega_0/Q$, where $\omega_0$ is the resonance frequency) play a crucial role in modern photonics and quantum optics[9,15], facilitating extreme temporal coherence[16], with lifetime on the time scale of $\sim Q/\omega_0$. Numerous essential applications, including optical frequency combs[1,2], monochromatic lasers[3,4], low-photon-number nonlinear optics[5], unidirectional nano-emitters[6], and cavity quantum electrodynamics studies[7,8], critically depend on high-$Q$ resonators. This has led to extensive efforts in designing ultrahigh-$Q$ resonators across different spectral regimes. However, experimentally realizing very large $Q$-factors at shorter wavelengths, e.g., visible wavelengths, remains an outstanding challenge due to material and fabrication constraints. The primary difficulty lies in fabrication imperfections that cause unwanted scattering loss and deviations from the intended design[9–12], which are much more severe in visible-wavelength devices because their small feature sizes are on a similar scale as fabrication defects.

At visible wavelengths, million-scale ($10^6$) $Q$-factors have been accessed in integrated photonic resonators, such as micro-ring and micro-disk cavities[9,17]. Due to tight spatial confinement of light in integrated photonics, one can minimize the defect-sensitive area in these resonators. In contrast, free-space lattice-resonant implementations such as metasurfaces and photonic crystal slabs have larger functional area (which enables large space-bandwidth product), and thus are highly defect-sensitive and face additional challenges, including long-range non-uniformity, substrate-induced out-of-plane asymmetry, and modal dispersion. Experimentally reported $Q$-factors in these free-space resonators typically fall within a scale of only $10^3$ (see Extended Data Table 1)[18–23] in the visible regime.

Despite these challenges, there is an outstanding and ever-growing demand for ultrahigh-$Q$ free-space optics due to their distinct advantages of large space-bandwidth product, easy free-space access, and parallel signal/data operations[13,14]. These unmet needs have spurred research into topological metasurfaces that are meticulously engineered to be more resilient to fabrication defects[24–26], for instance, by merging multiple bound states in the continuum (BICs) in momentum space[24]. However, in these approaches, the substrate needs to be removed to minimize out-of-plane asymmetry and the lack of lossless high-index materials at visible wavelengths may also limit design feasibility. Moreover, despite the improvements, the experimental $Q$-factors still fall short by orders of magnitude compared to those in integrated-optics resonators.

Here we design and experimentally demonstrate a million-scale ultrahigh-$Q$ guided mode resonance (GMR) at visible wavelengths in a resist-based etch-free metasurface (Fig. 1). Etching is the major process that introduces roughness and defects. An 'etch-free' design procedure aims to minimize fabrication imperfections, rather than engineering the device's robustness to imperfections. The advantage of this novel concept was recently validated by Huang *et al.* in a near-infrared ultrahigh-$Q$ metasurface[27] and by Ko *et al.* in a mid-infrared ultrahigh-$Q$ microresonator[28], among others[29,30]. Expanding on these findings, we adapt the etch-free metasurface strategies[27,29,30] under the material constraints of visible regime, and advance visible-wavelength free space meta-optics into a new era of million-$Q$ performance. We achieve this by using a perturbed multilayer-waveguide configuration as shown in Fig. 1a with material refractive indices not exceeding 2.0 (SiN). We emphasize that low-loss high-index materials are rare at visible wavelengths, and hence our moderate-index design facilitates generalization across different platforms.

To characterize a million-$Q$ resonance under free-space visible wavelength excitations, we develop a new laser-scanning momentum-space-resolved spectroscopy technique. It combines a tunable-wavelength laser and a momentum-space imaging system to visualize the modal dispersion[24,31–34] in the full energy-momentum space with ultrahigh resolution in wavelength (~0.42 pm) and angle (~0.028°).

Our platform can be readily used in various applications including optical sensing, filtering, quantum light sources, and few-photon nonlinear optics. As an example, we integrate monolayer WSe$_2$ into our ultrahigh-$Q$ meta-optical resonator and demonstrate highly unidirectional and narrow-linewidth exciton emission. The ultrahigh-$Q$ cavity effectively boosts the density of states at the Γ point, and a pump-power-dependent emission concentration towards the Γ point under continuous-wave (CW) laser pumping is observed at room temperature.

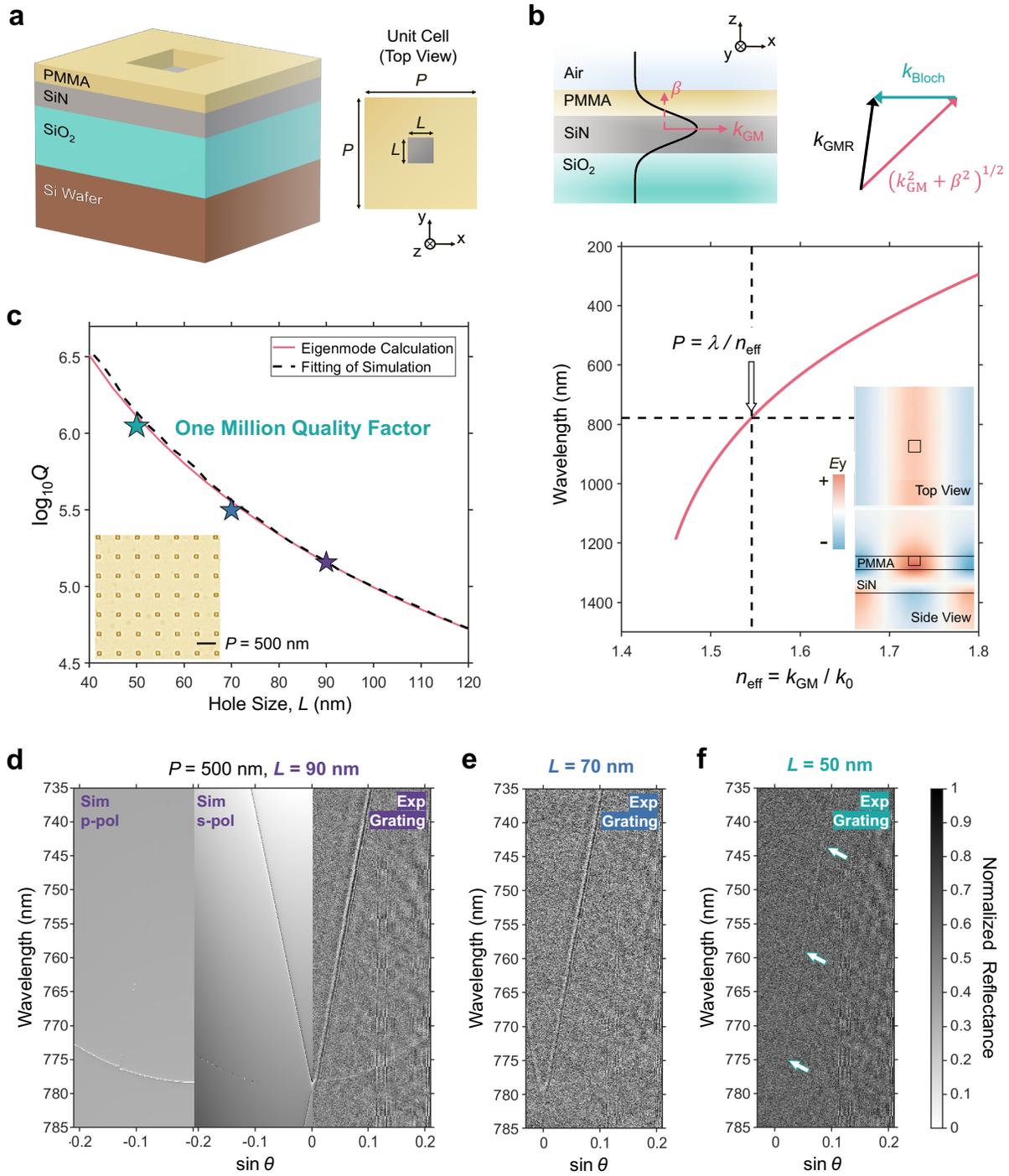

**Fig. 1. Ultrahigh-$Q$ guided mode resonances (GMRs) in an etch-free metasurface at visible wavelengths.** (**a**) Schematic showing the unit cell of the etch-free metasurface. From top to bottom, the geometry includes a 58-nm-thick patterned PMMA layer, a 100-nm-thick SiN layer, and a 1470-nm-thick SiO$_2$ layer on a Si substrate. The patterning is defined by the period $P$ and defect hole size $L$. (**b**) Design principle, from guided modes to ultrahigh-$Q$ GMRs: Top-left, Schematic of an air/58-nm-PMMA/100-nm-

SiN/SiO$_2$ multilayer slab waveguide; Top-right, Wave vector analysis when a periodic perturbation is patterned at the PMMA layer. The Bloch momentum $k_{Bloch} = 2\pi/P$ introduced by periodic patterning couples guided modes into free space. $k_{GM}$ and $\beta$ are the in-plane and out-of-plane components of the guided mode wave vector, respectively. $k_{GMR}$ is the GMR wave vector; Bottom, Dispersion of the TE$_1$ guided mode in the studied system, which guides the design of GMR. The dashed black lines show an example of designing a Γ-point resonance at 779 nm. The insets are the corresponding 779 nm GMR mode profiles when $P$ = 500 nm and $L$ = 50 nm. (**c**) $Q$-factor of GMR as a function of $L$ when $P$ is fixed as 500 nm. The red curve is predicted by an eigenmode solver in COMSOL Multiphysics®. The dashed black curve comes from the Fano fitting of the simulated reflectance spectra from Lumerical RCWA. The experimental results using the same fitting method are highlighted as three stars, up to one million $Q$. The inset is an SEM image of the fabricated metasurface under top view. False colors are added to highlight the patterned PMMA. Scale bar, 500 nm. (**d-f**) Simulated and experimentally measured momentum-space-resolved reflectance spectra of our meta-resonators. With a fixed $P$ of 500 nm, devices with different $L$ are studied: (d) $L$ = 90 nm, (e) $L$ = 70 nm, (f) $L$ = 50 nm. A spectrometer with a 1200-lines/mm grating is used to differentiate the wavelength in spectra. The arrows in (f) guide the eyes towards the GMR feature of interest that is vague due to the ultranarrow linewidth beyond the wavelength resolution. A difference operation is performed, subtracting the background (at unpatterned-PMMA areas) signals from the device signals, to eliminate the multilayer interference influence. The spectra are normalized to their respective maxima (see Methods).

**Design of ultrahigh-$Q$ GMR meta-resonator**

The proposed metasurface configuration and the design flow from a nonradiative guided mode to a radiative ultrahigh-$Q$ GMR are illustrated in Figs. 1a, b, respectively. To begin, we consider the four-layer slab waveguide in Fig. 1b, which comprises, from top to bottom, of a semi-infinite air superstrate, a 58-nm-thick layer of polymethyl methacrylate (PMMA), a 100-nm-thick layer of SiN, and a semi-infinite SiO$_2$ substrate. This stack supports both transverse electric (TE) and transverse magnetic (TM) guided modes with different cut-off wavelengths. For simplicity, we target the wavelengths where only the fundamental TE modes exist (see Supplementary Note 1), for which the electric field is $\mathbf{E}(x,z) = \hat{\mathbf{y}} E_y(z) e^{ik_{GM}x}$. $k_{GM}$ is the propagation constant (i.e., in-plane component of the guided mode wave vector), which can be found by solving the transcendental dispersion equation (detailed derivation in Supplementary Note 1),

$$\tan \beta_3 h_{SiN} = \frac{\beta_3(\kappa_1\kappa_4 - \beta_2^2)\tan(\beta_2 h_{PMMA}) + \beta_3\beta_2(\kappa_1+\kappa_4)}{(\beta_3^2\kappa_1 + \beta_2^2\kappa_4)\tan(\beta_2 h_{PMMA}) + \beta_2(\beta_3^2 - \kappa_1\kappa_4)}. \qquad (1)$$

In this equation, $\beta_2 = [n_{PMMA}^2 k_0^2 - k_{GM}^2]^{1/2}$ and $\beta_3 = [n_{SiN}^2 k_0^2 - k_{GM}^2]^{1/2}$ are the out-of-plane wave vector components in the PMMA and SiN layers, respectively, where $k_0 = 2\pi/\lambda$; $\kappa_1 = [k_{GM}^2 - n_{air}^2 k_0^2]^{1/2}$, and $\kappa_4 = [k_{GM}^2 - n_{SiO_2}^2 k_0^2]^{1/2}$ are the decay constants in air and SiO$_2$, respectively. The symbols $n$ and $h$ represent the refractive indices and thicknesses of different media. The plot in Fig. 1b shows the dispersion of the effective index $n_{eff} = k_{GM} / k_0$ of the TE$_1$ mode obtained by solving Eq. (1). In Supplementary Fig. S1, we also examine a simplified three-layer model that ignores the very thin resist layer[27], showing a significant deviation of $n_{eff}$ compared with the complete four-layer model. This suggests that the very thin PMMA plays a crucial role here, and the optical near field is expected to be effectively trapped in both the SiN and PMMA layers.

By creating a square periodic array of perturbations in the PMMA layer as shown in Fig. 1a, we introduce a structure-induced Bloch momentum $k_{Bloch} = 2\pi/P$ to compensate for $k_{GM}$ and thereby open a radiative leaky channel for the infinite-$Q$ guided mode to couple into free space. A wave vector analysis sketch is presented in the top-right panel of Fig. 1b. This leads to a GMR with wave vector $k_{GMR}$, whose resonance wavelength is determined by the period of perturbation $P$ and the guided mode dispersion, as shown in the bottom panel of Fig. 1b. For instance, to design a Γ-point (normal-incidence) GMR at a given wavelength $\lambda$, we need a $k_{GMR}$ with zero in-plane component, $n_{eff} \cdot \frac{2\pi}{\lambda} - \frac{2\pi}{P} = 0$. The dashed lines in the dispersion plot give an example of determining the value of $P$ for a resonance at 779 nm. Moreover, with a fixed $P$, we can use the guided mode dispersion curve to predict the GMR dispersion (Supplementary Fig. S2). The inset shows the simulated mode profiles of a 779 nm resonance at the Γ point when $P = 500$ nm and the defect hole size $L = 50$ nm. From the side view, the resonant near field is clearly trapped at the interface of SiN and PMMA, as predicted by the four-layer model. This feature can guide the design strategy to maximize the field overlap with the integrated functional materials, which will be further discussed in the last section.

Similar to a quasi-BIC leaky mode[26,35], the extent of periodic perturbations (hole size) applied on a photonic bound state (nonradiative guided mode) determines the $Q$-factor and radiative amplitude of the leaky mode (GMR)[27]. In our metasurface, a smaller $L$ induces a larger $Q$-factor as shown in Fig. 1c, accompanied by a decreased reflectivity amplitude as shown in Supplementary Fig. S3 (the amplitude decrease is negligible for the hole sizes chosen in our experimental demonstration). The predicted $Q$-factors in Fig. 1c are obtained through eigenmode simulations (red solid line) and Fano fitting of simulated reflection data (black dashed line). The close agreement between these methods validates the fitting approach, which is then applied to all experimental spectra (see Methods).

Three metasurfaces with fixed $P = 500$ nm and varying $L$ (90, 70, 50 nm) are fabricated and characterized. Note that, to ease the fabrication without affecting the design (Supplementary Fig. S4), the samples are fabricated on 1470-nm-$SiO_2$-on-Si wafers as illustrated in Fig. 1a. The $Q$-factors of these three devices (stars in Fig. 1c) all show excellent agreement with the theoretical expectation, yielding a record-high $Q$ of a million scale. Such good quantitative agreement is thanks to the reduced amount of fabrication imperfection enabled by the etch-free design, where only the PMMA layer is patterned by electron-beam lithography and a development process, eliminating the need for etching.

**Visualization of GMR dispersions**

Random device defects, environmental noise, light path misalignment, or even dust on optical components can cause 'ghost' narrow-linewidth spectral features in visible-wavelength free-space measurements. When the linewidth of the studied resonance mode is comparable to these high-frequency background signals, experimental spectral analysis becomes less convincing. To properly characterize an ultrahigh-$Q$ meta-resonator, we need momentum-space-resolved spectroscopy that enables the visualization of the complete modal dispersions, thereby distinguishing the real resonances from various noise sources[24,32,34]. Here, we start with a conventional energy-momentum reflectance spectroscopy that uses a white light source and a grating (see the optical setup in our previous publications[36,37]). As shown in Figs. 1d-f, the measured dispersion of all three devices matches well with the simulated GMR response in Fig. 1d, confirming the GMR nature of the measured high-$Q$ modes and again proving the near-ideal performance of our etch-free metasurfaces. Figure 1d also reveals the co-existence of different resonance modes with s- or p-polarization. A full eigenmode analysis can be found in Supplementary Note 2. In the main text, we focus on the s-polarized ultrahigh-$Q$ GMR with a clean linear dispersion, as highlighted by the arrows in Fig. 1f.

We find that, as the $Q$-factor increases with decreasing $L$ (from Fig. 1d to 1f), the ultrahigh-$Q$ GMR dispersion curve becomes faint. This suggests that the linewidth of the studied resonance mode is much smaller than the wavelength resolution (~0.05 nm) of the 1200-lines/mm grating in the setup. Since interference-induced high-$Q$ resonances typically manifest as an asymmetric Fano shape with one peak and one dip, insufficient wavelength resolution can average out the peak and dip, causing incorrect or missing spectral information.

To tackle this intriguing dilemma arising from a record-high $Q$ in visible-wavelength free-space optics — a unique and new task that could become common as our strategy can be easily

generalized — we introduce a laser-scanning momentum-space-resolved spectroscopy technique, as shown in Fig. 2a. Analogous to integrated photonic resonator measurements, a tunable laser is employed to scan the wavelength with extremely high resolution. Distinctively, the wavelength-scanning laser is integrated into a 4-*f* momentum-space imaging system, and a two-dimensional charge-coupled device (2D CCD) continuously captures the iso-frequency contours of the photonic band structures at different wavelengths determined by the tunable laser.

With full access to the multi-dimensional information in the wavelength/energy-momentum (*E-k*) space, we can visualize any *E-k* cross-section, such as a cut along $k_y$ in Figs. 2b-d. In most cases, we enable a small number of pixels on the 2D CCD, only collecting the desired *E-k* information, to save scanning time. See Methods and Supplementary Note 3 for technique details. As an aside, we target the 779-nm GMR designs for our experimental demonstrations because the only accessible tunable laser in our lab covers 765 to 781 nm. Devices working at shorter wavelengths have also been fabricated and studied (Supplementary Fig. S5). Ultrahigh-resolution quantitative characterization can readily be performed using the same method as long as a tunable laser is available at the relevant wavelength.

Figures 2b-d compare the results of conventional energy-momentum reflectance spectroscopy with our laser-scanning-integrated approach. Measured spectra from the metasurface with *P* = 500 nm and *L* = 90 nm are presented as an example. Dramatic improvements can be found in Figs. 2c, d (laser-scanning) compared to Fig. 2b (conventional), successfully detecting the ultrahigh-*Q* spectral feature in visible-wavelength free-space measurements.

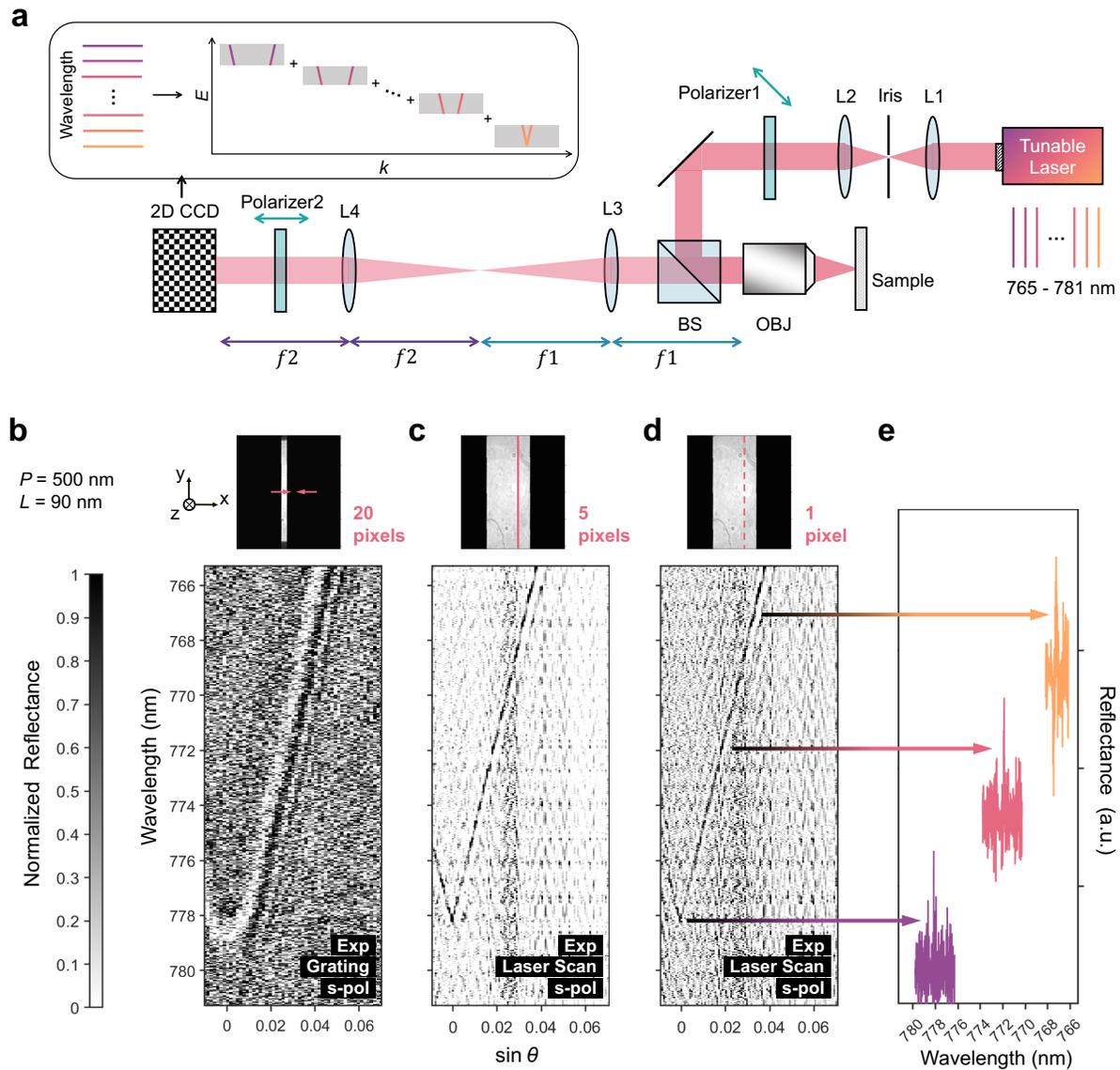

**Fig. 2. Laser-scanning momentum-space-resolved spectroscopy with ultrahigh resolutions in wavelength and angle of signal.** (**a**) Schematic of the optical setup and how laser-scanning defines wavelength in momentum-space-resolved spectroscopy. L1-4, lens. BS, beamsplitter. OBJ, objective. CCD, charged-coupled device camera. More details in optical setup and measurements can be found in Methods. (**b-d**) Comparison of the measured momentum-space-resolved reflectance spectra under different wavelength and angle resolutions. Spectra of the device with $P$ = 500 nm and $L$ = 90 nm are presented as an example. A spectrometer with a 1200-lines/mm grating provides a wavelength resolution of ~0.05 nm/pixel in (b), while a resolution of ~2.1 pm/pixel is achieved in (c, d) via a tunable laser. The angle resolution is ~0.076°/pixel in (b) with a 10X objective, and ~0.044°/pixel in (c, d) with a 2X objective.

Insets on top highlight the numbers of pixels summed along *X* axis on CCD when evaluating the momentum-dependent spectral response along *Y* axis. The more pixels summed, the more severe the dispersion-induced mode broadening is. (**e**) Reflectance spectra extracted at arbitrary *k* values in the momentum space from (d).

Besides improving the resolution in wavelength and momentum (angle), we also examine the influence of the number of pixels summed along the direction perpendicular to the visualized *E-k* cross-section (e.g., $\Delta k_x$ in *X* axis for a *E-k* cross-section cut along the *Y* axis). In common spectroscopy approaches using gratings, a slit must be applied as shown in the top inset of Fig. 2b. The slit width needs to balance the grating efficiency and resolution, and the optimal width in our setup turns out to be at least 20 pixels wide. This causes extra dispersion-induced mode broadening[37]. Excitingly, our technique allows to sum an arbitrary number of pixels, as illustrated in the top insets of Fig. 2c, d. With more pixels summed (Fig. 2c), the signal-to-background ratio is higher, while the single pixel case (Fig. 2d) can provide the narrowest GMR feature, closer to the intrinsic GMR property. In the following, we display the *E-k* spectra with a few pixels summed and extract *Q*-factors from single-pixel data.

**Direct measurement of a million *Q***

With sufficient resolutions in *E-k* space and a clear GMR dispersion picture, we can extract high-wavelength-resolution 1D spectra at arbitrary *k* values and confidently identify the high-*Q* peaks originating from the GMR meta-resonator, as shown in Fig. 2e.

Based on this, we fit a Fano lineshape to the extracted 1D spectra at the Γ point (*k* = 0, highlighted by the arrow in the right panel of Fig. 3a) and determine the *Q*-factors of three fabricated metasurfaces to be $1.10 \pm 0.09$ million (*L* = 50 nm, Fig. 3a), $313 \pm 8$ thousand (*L* = 70 nm, Fig. 3b), and 144 ± 5 thousand (*L* = 90 nm, Fig. 3c), respectively. As concluded in Extended Data Table 1, our report of a million *Q* in the experiment is orders of magnitude higher than state-of-the-art demonstrations in visible-wavelength free space meta-optics[18–23].

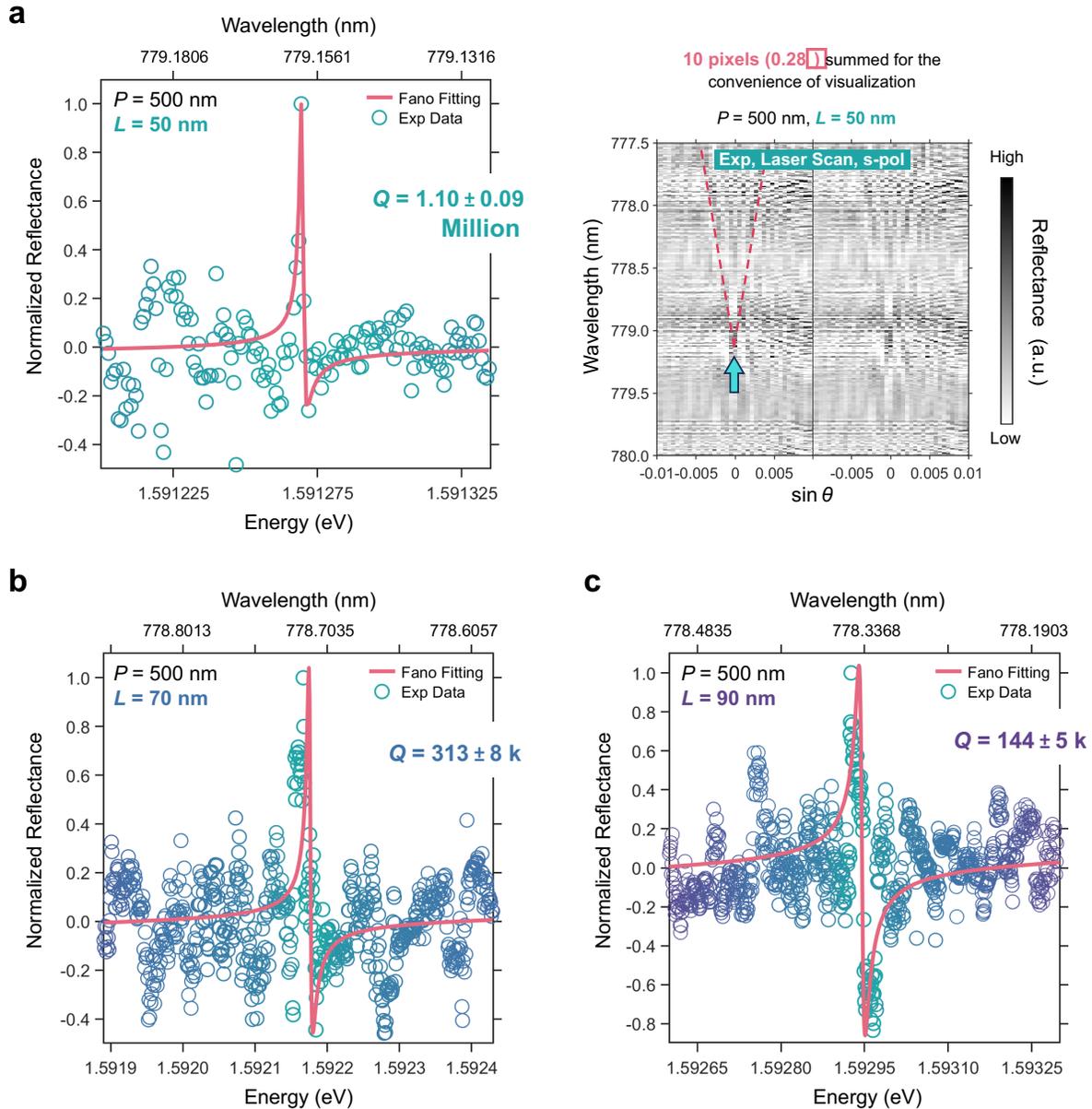

**Fig. 3. Experimental demonstration of million-*Q* GMR at visible wavelengths.** (**a**) Left, Reflectance spectra of the device with *P* = 500 nm and *L* = 50 nm at Γ point (normal incidence). The Fano fitting (pink curve) reveals a *Q* factor of 1.10 million. The Γ-point data are extracted from extra-fine laser-scanning

momentum-space-resolved reflectance spectroscopy with a wavelength resolution of ~0.42 pm/pixel and an angle resolution of ~0.028°/pixel; Right, the momentum-space-resolved spectra where the Γ-point data is extracted from. Note that a 0.28° range (10 pixels) of data are summed to broaden the ultranarrow mode linewidth for the convenience of visualization here, but only single-pixel data are used for Fano fittings. The same momentum-space-resolved spectra are plotted twice, one with the pink dashed lines highlighting the dispersion of GMR. The arrow points to the Γ-point resonance. (**b**, **c**) The same as (a) but for the devices with (b) $L = 70$ nm and (c) $L = 90$ nm. The fitted Q factors are 313 and 144 thousand respectively.

Looking beyond the million-$Q$ achievements, further advancements will require a faster 2D camera and a more stable tunable laser. Here, to obtain enough data points for accurately fitting a million-$Q$ resonance, we reduce the laser scanning speed while the 2D CCD capture frequency is already at its maximum, achieving a wavelength resolution as fine as ~0.42 pm. However, as shown in the right panel of Fig. 3a, an unwanted stripe-like background emerges in this extra-fine momentum-space-resolved spectrum, resulting from laser instability over the longer scanning time.

**A unidirectional and narrow-linewidth exciton emitter**

In this last section, we demonstrate the practical advantages of our ultrahigh-$Q$ meta-optical platform by integrating monolayer WSe$_2$ into the metasurface and showcasing a highly unidirectional and narrow-linewidth exciton emitter at room temperature. Monolayer transition metal dichalcogenides such as WSe$_2$ are promising next-generation nano-emitters but suffer from poor emission directionality[38] and broad spectral linewidth at room temperature[39]. As shown in the left part of Fig. 4a (white-to-purple colormap), the room temperature photoluminescence (PL) of a monolayer WSe$_2$ (under a 532nm CW laser pumping) in our unpatterned multilayer structure has a full-width-at-half-maximum (FWHM) spectral linewidth ($\varGamma_\lambda$) of ~25.12 nm and a FWHM linewidth in far-field emission angle ($\varGamma_\theta$) of ~120.95° (see simulated 3D emission directionality in Supplementary Fig. S6). Our ultrahigh-$Q$ platform can substantially narrow both of these important figures of merit by more than 95% (Fig. 4b).

As shown in the right part of Fig. 4a (grayscale), we select a metasurface period $P$ of 480 nm to match the Γ point of GMR with the WSe$_2$ PL in wavelength. A defect hole size $L$ of 70 nm is used, yielding a $Q$-factor of ~297 thousand, which is still higher than any reported $Q$ in literatures[18–23]. As illustrated in Fig. 4b, monolayer WSe$_2$ is placed between SiN and PMMA layers for better field overlap (see the mode profile inset in Fig. 1b). This is easily achieved by first transferring the functional material, i.e., exfoliated WSe$_2$, onto SiN and then performing the spin coating and lithography in PMMA.

The WSe$_2$ PL linewidths in the unpatterned structure and patterned ultrahigh-$Q$ meta-resonator are compared in Fig. 4b, in both spectral (left panel) and angular (right panel) dimensions, showing laser-like monochromaticity and unidirectionality when coupled with the metasurface at room temperature, albeit without any operating power density threshold. The applied CW pumping power density is as low as 0.555 kW/cm$^2$. The spectral and angular plots in Fig. 4b are extracted from the energy-momentum PL spectra of the WSe$_2$-metasurface hybrid in Supplementary Fig. S7, which reveal the highly concentrated emission at the Γ point of GMR.

Additionally, as we increase the 532 nm pump power density, we observe obvious PL narrowing in both spectral ($\Gamma_\lambda$) and angular ($\Gamma_\theta$) dimensions. A power-dependent analysis is performed, and the extracted crucial parameters are presented in Figs. 4c-e. It should be emphasized that the heating effect is carefully excluded in our CW laser pumping tests. Referring to a literature[40] that studied 532 nm laser-induced heating on WSe$_2$ (the same laser wavelength and material) with rigorous experiments and simulations, we set our CW laser power density to a safe range where no detectable temperature increase was found[40].

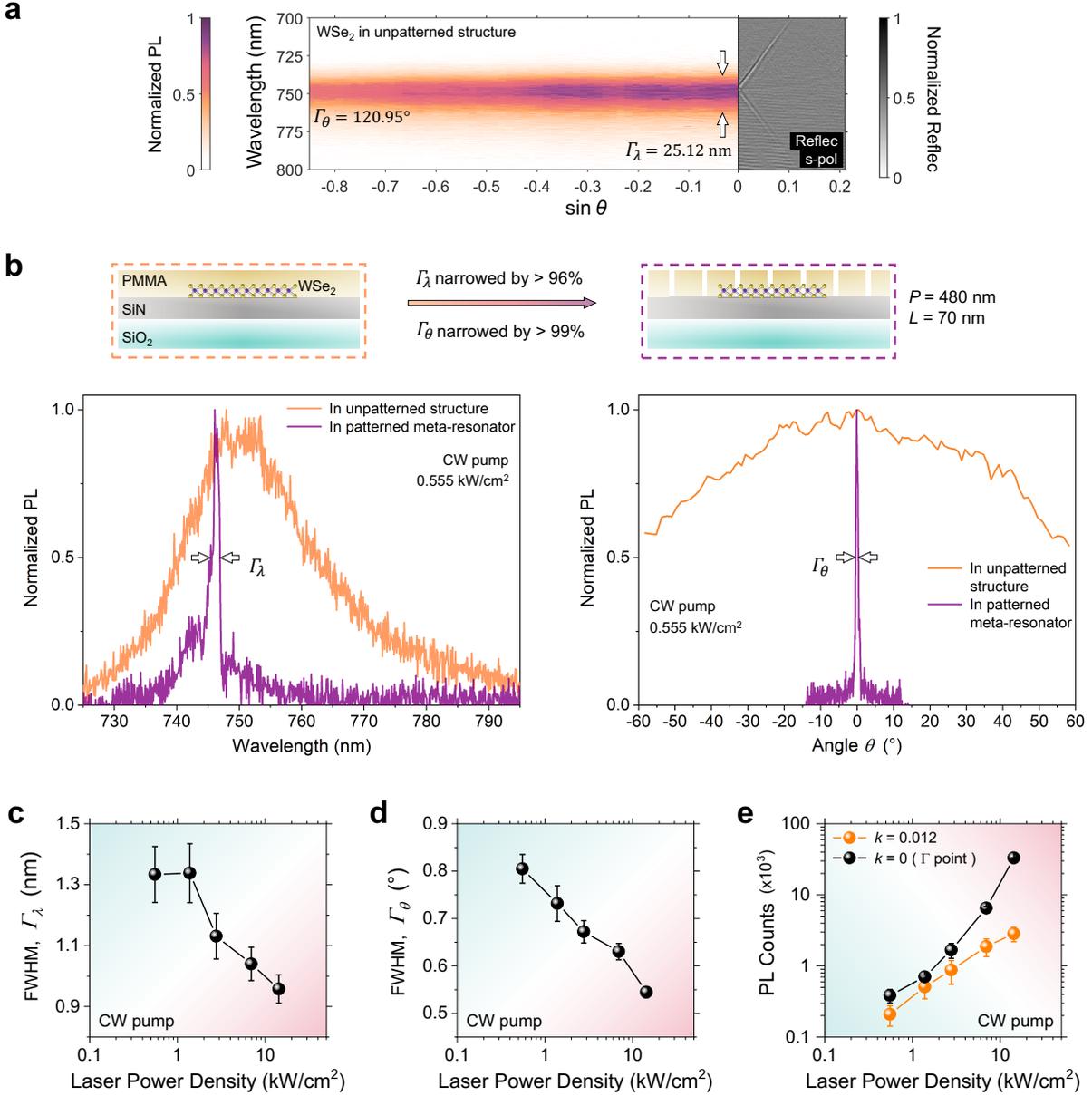

**Fig. 4. Highly unidirectional and narrow-linewidth exciton emission from monolayer WSe$_2$ when coupled with the ultrahigh-$Q$ meta-resonator at room temperature.** (**a**) Left, Energy-momentum photoluminescence (PL) spectra of monolayer WSe$_2$ in the unpatterned structure. Pumped by a 532 nm 0.555 kW/cm$^2$ CW laser. The broad full-width-at-half-maximum (FWHM) linewidths in far-field emission angle ($\Gamma_\theta$) and spectral wavelength ($\Gamma_\lambda$) are highlighted; Right, Energy-momentum reflectance spectra of the WSe$_2$-metasurface hybrid system, where $P$ = 480 nm and $L$ = 70 nm. (**b**) Narrowing the WSe$_2$ PL in both spectral and angular dimensions. Top, Schematics of monolayer WSe$_2$ in the unpatterned structure and patterned meta-resonator on a fused silica substrate (no Si wafer), respectively. The same device is measured before and after PMMA patterning. Bottom, Comparison of WSe$_2$ PL linewidths before and after patterning, in both spectral (Left) and angular (Right) dimensions. (**c-e**) Extracted emission properties of the WSe$_2$-metasurface hybrid from the energy-momentum PL spectra (Supplementary Fig. S7) as a function of pump power density (logarithmic scale): (c) spectral linewidth, FWHM $\Gamma_\lambda$, (d) angular linewidth, FWHM $\Gamma_\theta$, and (h) integrated PL counts (logarithmic scale) at $\sin\theta = 0.012$ (orange dots) and $\sin\theta = 0$ (black dots), respectively.

As shown in Figs. 4c, d, the FWHM $\Gamma_\lambda$ is narrowed from 1.334 nm to 0.957 nm (by ~28%), and FWHM $\Gamma_\theta$ is narrowed from 0.805° to 0.545° (by ~32%) under 14.33 kW/cm$^2$ CW pumping. To better understand this narrowing phenomenon, in Fig. 4e, we examine the integrated PL counts at $\Gamma$ point ($k = 0$) and at $k = 0.012$, slightly deviated from the $\Gamma$ point, as a function of pump power density. We choose a very small deviation from the $\Gamma$ point because the emission is already well concentrated even before we increase the pump power and PL signals at larger $k$ values are near zero. As shown in Fig. 4e, a nonlinear increase of PL counts at $k = 0$ is found, while there is a saturation trend in term of PL counts at $k = 0.012$, suggesting amplified spontaneous emission (ASE) in our system because of the extremely high density of state at the $\Gamma$ point.

We have also conducted PL studies at higher power densities by using a 532 nm 100-picosecond pulsed laser and a consistent power-dependent narrowing tendency is found (Supplementary Fig. S8), further supporting the ASE assumption. The narrowed FWHM $\Gamma_\lambda$ and $\Gamma_\theta$ eventually saturate at 0.587 nm and 0.269° under pulsed pumping. More discussions can be found in Supplementary Note 4.

Lastly, we notice strong in-plane scattering at the boundaries of the finite-size WSe$_2$ flake (see Supplementary Note 4 and Figs. S7, S10), possibly because of the significant refractive index difference between WSe$_2$ ($n > 4.4$) and our multilayer waveguide structure ($n < 2.0$). Consequently, we hypothesize that both the strong temporal confinement of our ultrahigh-$Q$ GMR and the considerable spatial confinement due to the high-index WSe$_2$ monolayer disk (acting like an in-plane resonator) contribute to the observed extreme emission concentration at RT and under CW pumping. Further studies are inspired but beyond the scope of this work.

**Summary**

In conclusion, we have experimentally demonstrated a record-high million-scale ultrahigh-$Q$ GMR at visible wavelengths under free-space excitation using an etch-free metasurface. Advancing free space meta-optics into a new era of million-$Q$ performance, we have also developed a laser-scanning momentum-space-resolved spectroscopy technique to accurately characterize arbitrary dispersive ultrahigh-$Q$ modes in the challenging free-space optical measurements. With material refractive indices not exceeding 2 in the demonstrated million-$Q$ device and the simple resist-based fabrication procedure, our strategy can be easily generalized across various platforms at visible wavelengths, inspiring more complicated meta-resonator designs (chiral[41], topological[24-27], nonreciprocal[42], reconfigurable[43], etc.) and functional material integration.

Furthermore, we have integrated monolayer WSe$_2$ into our ultrahigh-$Q$ meta-optical resonator and demonstrated a highly unidirectional and narrow-linewidth exciton emitter operating at RT and under CW pumping. Extreme PL concentration with FWHM linewidths as narrow as 0.269° in far-field emission angle and 0.587 nm in wavelength has been achieved, offering a laser-like light source without any power density threshold requirement. Pump-power-dependent linewidth narrowing and nonlinear increase in PL counts are observed, suggesting ASE in our system and highlighting the potential of our ultrahigh-$Q$ meta-optics platform for studying cavity-modulated quantum emitters and low-photon-number nonlinear optics. Additionally, our result holds great promise for applications like optical sensing, filtering, and metrology.

## Methods

### Metasurface fabrication

All metasurfaces without WSe$_2$ emitter integration (Figs. 1-3 and Supplementary Fig. S5) are fabricated on a 1470-nm-thick SiO$_2$ on silicon substrate (UniversityWafer, Inc.). The device for WSe$_2$ integration and PL measurements (Fig. 4) is fabricated on a JGS1 fused silica substrate (UniversityWafer, Inc.). A 100-nm-thick layer of Si$_3$N$_4$ is grown on top of the substrates through low-pressure chemical vapor deposition (LPCVD) by UniversityWafer, Inc. The wafer is then diced into 8×8 mm$^2$ chips (DAD321, Disco America).

    The chips are thoroughly cleaned by sonication in acetone, followed by isopropyl alcohol (IPA), each for 5 minutes. After that, the SiN surface is treated by oxygen plasma at 150 W for 5 minutes to dry the surface and remove any solvent residue (AutoGlow, Glow Research). Subsequently, a 58-nm-thick layer of positive-tone resist PMMA is spin-coated and annealed under 180 ºC for 3 minutes. After cooling the chip to room temperature, a layer of conductive polymer (DisCharge H$_2$O) is spin-coated on top. The metasurface pattern is defined using a JEOLJBX-6300FS 100kV electron-beam lithography system, followed by development in a cold water/IPA mixture for 2 minutes. The fabrication is etch-less, ensuring minimized sidewall roughness of the patterns.

### Preparation of monolayer WSe$_2$

WSe$_2$ flakes are directly transferred onto the SiN surface using the standard tape exfoliation method. Monolayers are identified by checking the imaging contrast under optical microscope and confirmed using atomic force microscopy (AFM).

### Optical setup and measurement

Optical characterizations are performed using a custom-built optical setup (Fig. 2a). Light is introduced into the setup via a single-mode fiber. The tunable laser (Newport TLB-6712), continuous-wave pump (Laserglow Technologies 532 nm DPSS Laser), and pulsed pump (NKT SuperK FIU-15 Laser, 78 MHz repetition rate, with a SuperK SELECT tunable multi-channel filter) are all coupled in this manner. The light is first focused on a 75 μm iris using L1 (focal length, $f =$ 60 mm) and L2 ($f =$ 75 mm) to ensure a uniform Gaussian beam. An objective lens is used to both focus the light onto the sample and collect the reflection/PL signals. To obtain the momentum-space-resolved spectrum, a telescope consisting of lenses L3 ($f =$ 180 mm) and L4 ($f =$ 150 mm) is employed. Here, L3 and L4 are confocal, L3 is in focus with the back focal plane of the objective lens, and L4 is in focus with the spectrometer CCD (Princeton Instruments Isoplane 160 with PIXIS 400).

In the laser-scanning momentum-space-resolved reflectance measurements (Figs. 2c, d and Fig. 3), a 2X Mitutoyo Plan Apo Infinity Corrected Long WD Objective (numerical aperture, NA 0.055) is used. To improve the signal-to-noise ratio, a crossed-polarization measurement method[24,27] (Supplementary Note 3) is employed by inserting two linear polarizers into the excitation and collection light paths, respectively, as shown in Fig. 2a. The second polarizer also determines the polarization of the GMR modes. Note that there is no mirror between Polarizer1 and BS in the actual setup, and a plate beamsplitter is used instead of a cube beamsplitter to avoid extra interference caused by reflections at the optical interfaces. For signal collection, we use the 2D CCD in the commercial spectrometer with the grating disabled. The wavelength scanning of the tunable laser and the data capture by the CCD are synchronized and controlled by a Python script.

In the PL measurements (Fig. 4), a 10X Olympus PLN Objective (NA 0.25) is used. With a 650 nm short-pass filter in the excitation pathway and a 550 nm long-pass filter in the collection pathway, we filter the pump laser and ensure a clean PL signal. A single linear polarizer is inserted into the collection light path to ensure a clean s-polarized GMR signal. For signal collection, we use the commercial spectrometer with the 1200-lines/mm grating enabled.

Spectroscopy data processing

In the reflectance measurements (Figs. 1, 2, 3), both the device and background (the unpatterned-PMMA areas next to the device) spectra are collected — $R_{device}$ and $R_{BG}$. First, a difference operation is performed, $R_0 = R_{device} - R_{BG}$, to eliminate the multilayer interference influence. Then, low-frequency noise signals are filtered, as our ultrahigh-$Q$ GMR signal is super high-frequency. The noise involves light source fluctuation, photodetector background, and the interference introduced by optical components like beamsplitters and lenses (the interference is severe when a tunable laser with good coherence is applied). The noise filtering is achieved via a customized two-dimensional bandpass filter in MATLAB, which transforms the data to the Fourier space by fast Fourier transform and selects the frequency components within an anisotropic elliptical donut-shaped region. The cutoff frequency is carefully chosen such that only the relatively low-frequency noise is filtered without affecting the high-$Q$ signal. A linearly gradual cutoff is implemented to mitigate the 'ring' artifact due to abrupt frequency cutoff. The code is available on: https://github.com/charey6/2D-frequency-filter.git. Finally, the spectra are normalized to their respective maxima.

In the PL measurements (Fig. 4), after subtracting the setup background signals, the measured spectra are directly normalized to their respective maxima.

Fano fitting

Spectra fittings are performed using a commercial software MagicPlot. The Fano fitting equation[44] below are manually written into the software:

$$R_{Fano}(\lambda) = a\left[\frac{\left(b + {}^{2(\lambda-\lambda_0)}/_\Gamma\right)^2}{1 + \left({}^{2(\lambda-\lambda_0)}/_\Gamma\right)^2} c + (1-c)\right],$$

where $R_{Fano}$ is the spectrum to be fitted. $a$, $b$, $c$ are constant real numbers. $\Gamma$ and $\lambda_0$ are the resonance FWHM linewidth and center wavelength. The $Q$-factor is then determined by $Q = \lambda_0/\Gamma$. The fitting errors are evaluated through the residual standard deviations automatically generated by the software in the fitting process. In specific, the software uses iterative Levenberg-Marquardt nonlinear least squares curve fitting algorithm to find the minimum residual sum of squares. Then, the corresponding residual standard deviation is used to describe the root mean square of the error (over all the data points) for the fitted parameters.

Numerical simulations

The commercial software Ansys Lumerical is used to simulate the momentum-space-resolved reflectance spectra of the metasurfaces (Fig. 1d) using rigorous coupled-wave analysis method (RCWA). As confirmed by ellipsometry (J.A. Woollam M-2000), we consider a 58-nm-thick PMMA with a refractive index of 1.46, a 100-nm-thick SiN with a refractive index of 1.99, and a 1470-nm-thick $SiO_2$ with a refractive index of 1.44. Based on SEM measurements (Thermo Fisher Scientific Apreo 1), a slightly deviated device period $P$ of 504 nm is applied. Additionally, through AFM (Bruker Dimension Icon), we find that the defect hole depth is only ~40 nm, not completely penetrating the PMMA layer. This factor is also considered in the simulations.

Eigenmode analyses are conducted using a commercial finite element method solver, the wave optics module in COMSOL Multiphysics® 5.2 and 6.0. For each valid complex eigenfrequency $f = \Omega + i\cdot\Gamma/2$ found, where $\Omega$ and $\Gamma$ are respectively the resonance frequency and damping rate, the $Q$-factor is determined by $Q = \Omega/\Gamma$.

**Extended data**

**Extended Data Table 1.** Experimentally reported *Q*-factors in free-space optics at visible wavelengths.

| Ref. | Q | λ (nm) | Design / Structure | Device size (μm²) |
|---|---|---|---|---|
| 18 | 8,000 | 750 | GMR / SiO$_2$ grating on SiN waveguide | 10,000 x 15,000 |
| 19 | 391 | 860 | GMR / photoresist grating on HfO$_2$ waveguide | \ |
| 20 | 32,000 | 490 | GMR / SiN photonic crystal slab | 600 x 600 |
| 21 | 10,000 | 583 | Symmetry-protected BIC / SiN photonic crystal slab | 7,000 x 7,000 |
| 22 | 2750 | 825 | Symmetry-protected BIC / GaAs metasurface | 60 x 108 |
| 23 | 2750 | 717 | Resonance-trapped BIC / TiO$_2$ lattice on dielectric-covered mirror | 500 x 500 |
| **This work** | **1,100,000** | **779** | **GMR / patterned PMMA resist on SiN waveguide** | **900 x 900** |

## Acknowledgements

J.F. thanks Mr. Suichu Huang for the insightful discussion. Work at University of Washington was supported by National Science Foundation (NSF) Grant No. DMR-2019444 and NSF-2103673. V.M.M. was also supported by NSF Grant No. DMR-2019444. Part of this work was conducted at the Washington Nanofabrication Facility / Molecular Analysis Facility, a National Nanotechnology Coordinated Infrastructure (NNCI) site at the University of Washington with partial support from NSF via awards NNCI-1542101 and NNCI-2025489. E.M.R., S.A.M. and A.A. were supported by the Simons Foundation and AFOSR Foundation. K.Y. and Y.Z. acknowledge the financial support of the National Institute of General Medical Sciences of the National Institutes of Health (R01GM146962).

## Author contributions

J.F., R.C. and A.Majumdar conceived the idea. J.F. designed the meta-resonators and experiments. E.M.R., S.A.M. and A.A. developed the analytical theories. R.C. and A.Manna fabricated the devices. A.Manna and S.P. prepared the WSe$_2$. D.S., J.F., A.K., R.C., A.Manna and C.M. performed the experiments. J.F., E.M.R., K.Y. and Y.Z. conducted the numerical simulations. H.R., A.T. and R.C. did the morphology characterizations. J.F., R.C. and D.S. interpreted the results together with all the authors. J.F. wrote the manuscript with input from all the authors and supervised the project with A.Majumdar. J.F., R.C. and D.S. contributed equally to this work.

**Competing interests**

The authors declare no competing interests.